\documentclass[twoside,twocolumn,british,english,preprintnumbers]{revtex4}
\usepackage{amsmath}
\usepackage{amssymb}
\usepackage{graphicx}
\usepackage{epsfig,bbm}
\usepackage{babel}
\usepackage{color}

\def\be{\begin{equation}}
\def\ee{\end{equation}}
\def\bea{\begin{eqnarray}}
\def\eea{\end{eqnarray}}
\def\d{\mathrm d}

\begin{document}

\title{The Scale of Inflation in the Landscape}

\author{F. G. Pedro}

\author{A. Westphal}

\affiliation{Deutsches Elektronen-Synchrotron DESY, Theory Group, D-22603 Hamburg, Germany}

\begin{abstract}
We determine the frequency of regions of small-field inflation in the Wigner landscape as an approximation to random supergravities/type IIB flux compactifications. We show that small-field inflation occurs exponentially more often than large-field inflation The power of primordial gravitational waves from inflation is generically tied to the scale of inflation. For small-field models this is below observational reach. However, we find small-field inflation to be dominated by the highest inflationary energy scales compatible with a sub-Planckian field range. Hence, we expect a typical tensor-to-scalar ratio $r\sim {\cal O}(10^{-3})$ currently undetectable in upcoming CMB measurements.

\end{abstract}

\preprint{DESY-13-044}

\maketitle

\section{Introduction}
 
Recent years have seen both the advent of precision cosmology giving strong indications~\cite{Story:2012wx,Hinshaw:2012fq,Sievers:2013wk,Riess:2011yx,Riess:1998cb,Perlmutter:1998np,Blake:2011en,Padmanabhan:2012hf,Anderson:2012sa} for an early phase of cosmological inflation~\cite{Guth:1980zm,Linde:1981mu,Albrecht:1982wi,Linde:1983gd,Baumann:2009ds}, and theoretical evidence for an exponentially large landscape of meta-stable de Sitter vacua~\cite{Bousso:2000xa,Kachru:2003aw,Susskind:2003kw,Douglas:2003um,Denef:2004cf,Douglas:2006es} combined with the first models of inflation in string theory~\cite{Kachru:2003sx,Baumann:2009ni,Cicoli:2011zz}. As the number of inflationary model realizations and final states provided by dS vacua with small vacuum energy is quite possibly extremely large, a description of inflationary observables is in need for a statistical description if one wishes to move beyond the lamp posts given by existing model constructions.

Inflationary models are generically sensitive to the presence of higher-dimension operators (e.g. from radiative corrections or integrating out heavy fields), and this sensitivity naturally splits the model space into two parts~\cite{Baumann:2009ds}. In small-field models of inflation~\cite{Linde:1981mu,Albrecht:1982wi} the effective canonically normalized inflaton scalar field evolves parametrically less than a Planck distance in field space during the 60 efolds of cosmologically necessary inflationary expansion. Control of dimension-six corrections to the scalar potential is sufficient for this class. Large-field models~\cite{Linde:1983gd} involve the inflaton crossing a parametrically super-Planckian distance $\Delta\phi_{60}$ during the same 60 efolds. In such models, successful slow-roll inflation necessitates the suppression of corrections at any dimension which amounts to the presence of a protecting symmetry~\cite{Baumann:2009ds}. 
The only extant symmetry capable of protecting large-field inflation and which has been embedded into string theory so far has been a shift symmetry of an axion-like pseudo-scalar field. These axions arise generically in string compactifications~\cite{Freese:1990rb,Banks:2003sx,Dimopoulos:2005ac,Svrcek:2006yi} where they can yield large-field inflation using monodromy~\cite{McAllister:2008hb}.

Generically, these two classes are accompanied by an observational discriminator. Inflation produces primordial curvature perturbations and gravitational waves with nearly scale-invariant power spectra ($\Delta_{\cal R}^2\sim H^2/\epsilon$, and $\Delta_{\cal T}^2\sim H^2$, respectively) originating as quantum fluctuations stretched to super-horizon wavelengths. The fractional power in gravity waves (tensor modes) $r = \Delta_{\cal T}^2 / \Delta_{\cal R}^2=16\epsilon$ is controlled by the first slow-roll parameter $\epsilon= {\cal L}_{kin}/2H^2\ll 1$. Its smallness enforces a vacuum-energy like equation of state during inflation which is necessary to drive accelerated expansion. For a large class of models the slow-roll of the inflaton translates into a monotonically increasing evolution of $\epsilon$. This leads to a relation between $\Delta\phi_{60}$ and the scale of inflation $H$ which implies that large-field inflation is necessary to produce a sizable tensor mode fraction $r\gtrsim 0.01$ in reach technologically during the next few years~\cite{Lyth:1996im}.\footnote{Exceptions to this generic situation can arise, if the scalar potential is tuned to avoid a monotonic evolution of $\epsilon$~\cite{BenDayan:2009kv,Hotchkiss:2011gz}, if a second scalar field provides additional vacuum energy as in hybrid inflation~\cite{Baumann:2009ds}, or if there are additional light degrees of freedom coupled to the inflaton whose quantum vacuum fluctuations can convert back into additional tensor modes if properly arranged~\cite{Senatore:2011sp,Barnaby:2012xt,Kobayashi:2013awa}.}

By being tied to the scale of inflation, the tensor mode fraction $r$ is an inflationary observable which will at most have a statistical description on the landscape. Hence, we need to determine the distribution of inflationary vacuum energies for accessible regions of the landscape. A guiding motivation here is that an analysis of the distribution of extremely small vacuum energies close to zero on the landscape has already been successful in providing an anthropic explanation of the smallness of the observed positive late-time cosmological constant (c.c.)~\cite{Weinberg:1987dv,Bousso:2000xa}. The vacuum energy distribution very roughly factors into a contribution coming from a number count of inflationary solutions, and a cosmological factor which involves vacuum transitions described by tunneling events~\cite{Coleman:1980aw} and the subtleties of eternal inflation. 

Recent work has analyzed the cosmological probability distribution factor~\cite{Westphal:2012up}. This led to the surprising answer that the physics of tunneling-mediated vacuum transitions and eternal inflation largely decouple from the distribution of vacuum energies parametrically smaller than the Planck density. Hence, the cosmological prior is flat which leaves the inflationary vacuum energy distribution on the landscape to be determined to leading order by model realization and vacuum counting. We are thus left with comparing the relative number frequencies of small-field and large-field inflation models on an accessible region of the landscape which we here choose to be the landscape of type IIB flux compactifications on warped Calabi-Yau manifolds (CYs).

Hence, in this note we determine the number frequency count of small-field inflation models on the landscape of supersymmetric type IIB CY flux vacua.
Using random matrix theory, we find that there are exponentially many more small-field inflation models in the moduli potential of the type IIB flux landscape than there are proper dS vacua. Comparing this with the restrictions on large-field models occurring on this landscape discussed in~\cite{Westphal:2012up}, we therefore statistically expect the absence of primordial tensor modes $r\gtrsim 0.01$ in upcoming CMB observations.

  
\section{The Wigner ensemble and Random Supergravities}

The F-term potential of $\mathcal{N}=1$ supergravity
\be
V=e^{K}\left(F_A \bar{F}^A-3 |W|^2\right)
\label{eq:VF}
\ee
is the starting point of the analysis of critical points in the landscape. As usual $F_A=\partial_A W + W \partial_A K$ and $W$ and $K$ are the superpotential and the K\"ahler potential respectively. Critical points are defined by the condition 
\be
\partial_A V |_{cp}=0
\ee
and can be maxima, minima or saddles. To determine the nature of a given critical point one must analyse the eigenvalues of the Hessian matrix, defined in terms of the F-term potential as $\mathcal{H}_{m n}=\partial_{m n} V$ where $m, n$ can be holomorphic or anti-holomorphic indices. Taking into account the structure of the F-term potential of Eq. (\ref{eq:VF}), the Hessian decomposes into a sum of the form
\be
\mathcal{H}=\underbrace{\mathcal{H}_{SUSY}+\mathcal{H}_{K^{(3)}}}_{Wishart+Wishart}+\underbrace{\mathcal{H}_{pure}+\mathcal{H}_{K^{(4)}}}_{Wigner}+\mathcal{H}_{shift}.
\label{eq:H}
\ee
Each of these matrices is defined in terms of the K\"ahler potential, the superpotential and their derivatives \cite{Denef:2004cf,Marsh:2011aa}. For our purposes it suffices to review the definitions and some properties of the  Wishart and Wigner matrices (for a review see \cite{Metha:1967,Guhr:1997ve,Edelman:2005}).

A Wishart matrix \cite{Wishart:1928} is a complex matrix  defined as $M=A A^\dagger$
where A is a random $N_f\times N_f$ complex matrix drawn from some distribution with mean $\mu$ and variance $\sigma$: $\Omega(\mu,\sigma)$.
Its eigenvalue spectrum has support on the interval $[0,4 N_f\sigma^2[$, is peaked towards the origin and is given by the Marcenko-Pastur law \cite{Marchenko:1967}.

A Wigner matrix is a Hermitian matrix defined as $M=A+A^\dagger$, where $A$ is drawn from a distribution $\Omega(\mu,\sigma)$. The eigenvalue spectrum of the Wigner ensemble is given by the Wigner semi-circle law 
\be
\rho(\lambda)=\frac{1}{2\pi N_f \sigma^2}\sqrt{4 N_f \sigma^2-\lambda^2}
\label{eq:semicircle}
\ee
which can be obtained by unconstrained integration of the joint probability density function (pdf)
\be
dP(\lambda_1,...,\lambda_{N_f})= exp\left( -\frac{1}{\sigma^2} \sum_{i=1}^{N_f} \lambda_i^2\right)\prod_{i<j}(\lambda_i-\lambda_j)^2
\label{eq:jointpdf}
\ee
over all but one variable. Equation (\ref{eq:jointpdf}) gives the probability of generating a matrix with eigenvalues in $[\lambda_i,\lambda_i+\delta \lambda]$ and it will be crucial for the analysis of the probability of inflation in the landscape of random supergravities we will present later.  A rather useful physical interpretation of Eq. (\ref{eq:jointpdf}) was put forward by Dyson in \cite{Dyson:1962es} in terms of a one dimensional gas of charged particles moving under the influence of an attractive quadratic potential and a repulsive mutual interaction. This picture proves very useful in  qualitatively estimating behaviour of the system. 

 A crucial property of the eigenvalue spectrum of the Wigner ensemble is that for the cases of interest, in which the random matrices are drawn form a distribution $\Omega(0,1/\sqrt{2 N_f})$, it has support on the interval $[-\sqrt{2}, \sqrt{2}]~M_P^2$. So unlike the Wishart ensemble, which has all eigenvalues positive, a typical $N_f\times N_f$ matrix in the Wigner ensemble will have $N_f/2$ tachyonic directions.

The typical eigenvalue spectrum of random supergravities, as defined by  $\mathcal{H}$, was found analytically in \cite{Marsh:2011aa} through the free convolution of the constituent spectra. The spectrum has support in $\sim[-0.7, 7.5] ~M_P^2$ (for Minkowski vacua) and so it typically features several tachyonic directions, meaning that the most likely critical points in random supergravity are steep saddles rather than a local minima. 

While the eigenvalue spectrum of the full random supergravity is distinct from that of a Wigner matrix, it is certainly true that its tachyonic part has its origin in the Wigner matrix since the spectrum of the sum of Wishart matrices is positive definite.

The presence of the positive semi-definite contribution from the Wishart matrices in the full random supergravity leads to a substantially enhanced frequency of local minima compared to a Wigner matrix based estimated. However, as the frequency of inflationary regions relative to local minima is dominated by the tachyonic part of the spectrum originating in the Wigner matrix spectrum alone, this relative likelihood of inflation is still determined to leading order by the Wigner matrix estimate in the full random supergravity as well. Conversely, the absolute frequency of inflationary regions will be enhanced in the full random supergravity proportional to the increased occurrence of local minima.

Studies of the string landscape often involve computation of the probability of occurrence of critical points, with particular emphasis on minima, suited for description of the present day Universe. These spectra correspond a large the shift of the smallest eigenvalue to the right of its typical position and are exponentially unlikely \cite{Aazami:2005jf,Dean:2006wk,Marsh:2011aa}:
\be
P_{min}\sim e^{-c N_f^p + \mathcal{O}(N)}\qquad p\sim \mathcal{O}(1).
\label{eq:Pmin}
\ee

In this letter we analyse small field inflation in the same light and try to determine how likely it is to find sufficiently flat saddle points in the landscape using the Wigner ensemble as our main tool. The reasons to approximate the full Hessian by a single Wigner matrix are twofold: firstly it is the Wigner matrix that gives rise to the tachyonic directions and so by focusing on these  one hopes to uncover the inflationary structure behind the full Hessian; secondly for the Wigner ensemble we are in possession of the joint pdf,  Eq. (\ref{eq:jointpdf}), whose numerical integration allows us to estimate probabilities without recurring to direct counting. The joint pdf that lies behind the full Hessian of random supergravities, Eq. (\ref{eq:H}), is unknown and so direct counting, the generation of large samples of matrices and the counting of the ones that have the spectra we are looking for,  is the only probe available. Since we are looking for minima and flat saddle points, which are extremely rare events, direct counting is computationally expensive.

We therefore focus our analysis on the Wigner ensemble, presenting the results in the next section.

\section{Inflation in the Landscape}

We start by deriving an identity regarding the probability for inflation in the Wigner landscape. As explained above, the distribution of saddle points in a random supergravity will be given by the Wigner ensemble as the leading approximation to the full supergravity Hessian. By simple manipulation of the integration limits it is possible to prove that inflationary saddle points are exponentially more abundant than minima with masses greater than the inflationary mass. For our purposes, $q$-field inflation happens in a saddle point in which $q$ fields have masses in the range $[-\eta,\eta]$ and $N_f -q$ fields in  $[\eta,\infty[$, for suitably small $\eta>0$.

The probability for generating a Wigner matrix with all eigenvalues greater than $-\eta$ can be found by integration of the joint pdf: 
\be
\begin{split}
P&(\forall \lambda> -\eta)=\prod_{i=1}^{N_f}\, \int\limits_{-\eta}^\infty\d\lambda_i \d P(\lambda_1,...,\lambda_{N_f})\\
&=\sum_{n=0}^{N_f}\frac{N_f!}{n!(N_f-n)!} \prod_{i=1}^{n}\,\int\limits_{-\eta}^\eta \d\lambda_i \prod_{j>n}^{N_f}\,\int\limits_\eta^\infty\d\lambda_j \d P.
\end{split}
\label{eq:pdfInt}
\ee
In going from the first to the second line of (\ref{eq:pdfInt}) we have simply split the integration region into $[-\eta,\infty[=[-\eta,\eta[\cup[\eta,\infty[$ for each $\lambda$, taking care to include the correct combinatorial factors.
Using Dean and Majumdar's result regarding the probability of large fluctuations of extreme eigenvalues for the Wigner ensemble \cite{Dean:2006wk}
\be
P(\forall \lambda> \xi)= e^{-2 \Phi(\xi) N_f^2},
\label{eq:FitExp}
\ee
where $\Phi(\xi)$ is given by
\be
\begin{split}
\Phi(\xi)=&\frac{1}{108}\left[36 \xi^2 -\xi^4 +(15 \xi +\xi^3)\sqrt{6+\xi^2}+\right.\\
&\left.+27\left(\log18 -2\log(-\xi+\sqrt{6+\xi^2})\right)\right],
\end{split}
\ee
one may write Eq. (\ref{eq:pdfInt}) as
\be
\frac{P(inf)}{P(\forall \lambda> \eta)}= e^{2 \Delta c N_f^2}-1,
\label{eq:analyticResult}
\ee
with $\Delta c\equiv \Phi(\eta) -\Phi(-\eta)$. Henceforth $P(inf)$ denotes the total probability for inflation, defined as the sum over all possible inflationary dynamics for a given $N_f$, i.e. 
\be
P(inf)=\sum_{q=1}^{N_f} P(q-inf),
\label{eq:inf}
\ee

In a manifestation that it is statistically more expensive to displace the lowest eigenvalue to $\eta$ than to $-\eta$, we see that $\Delta c>0$ and so flat saddle points, suited for inflation, are exponentially more frequent in the landscape than minima with all masses larger than $\eta$. 

The main aim of this work is to determine the ratio $P(inf)/P(min)$, where we define  $P(min)=P(\forall \lambda>0)$. Once again the results of  \cite{Dean:2006wk} allow us to push ahead. Noting that
\be
\frac{P(min)}{P(\forall \lambda >\eta)}= e^{-2 (\Phi(0)-\Phi(\eta))N_f^2}\equiv e^{-2\widetilde{\Delta c}N_f^2}
\ee
one finds 
\be
\frac{P(inf)}{P(min)}= (e^{2 \Delta c N_f^2}-1)e^{2 \widetilde{\Delta c} N_f^2}\sim e^{2 \eta \Phi'(0) N_f^2} +\mathcal{O}(\eta^2).
\ee
We therefore expect inflationary saddle points to be exponentially more abundant than local minima in the Wigner landscape.

In order to confirm and extend the above results we estimate the relevant probabilities by Montecarlo integration of Eq. (\ref{eq:jointpdf}), setting $\eta=0.1$, in  the window $N_f \in [2,16]$.  We then fit the relevant probabilities for each value of $N_f$ to the exponential law of Eq. (\ref{eq:FitExp})
as is expected from the theory of large eigenvalue fluctuations developed in \cite{Dean:2006wk}. The results are presented in table \ref{fitResults}. 
\begin{table}[h!]
\centering
\begin{tabular}{l|c | c}
  &Analytical & Fit \\
\hline
$P(\lambda > -\eta)$ & $0.447$ &$0.429\pm0.004$ \\
$P(min)$& $0.549$ & $0.530\pm0.004$ \\
$P(\lambda > \eta)$& $0.665$ &$0.645\pm0.004$ \\
$P(inf)$& -- &$0.403\pm0.002$ \\
\hline
\end{tabular}
\caption{Analytical estimates and fits to numerical data.}
\label{fitResults}
\end{table}
We see that our method systematically overestimates the probabilities of occurrence of these rare events. This is reflected on a shift of the fitted parameters on the level of a few percent. We stress that even though the error bars cannot account for this deviation, the fact that the numerical and analytical results show the same trend lends credibility to our results. 

\begin{figure}[h!]
\centering
\includegraphics[width=0.45\textwidth]{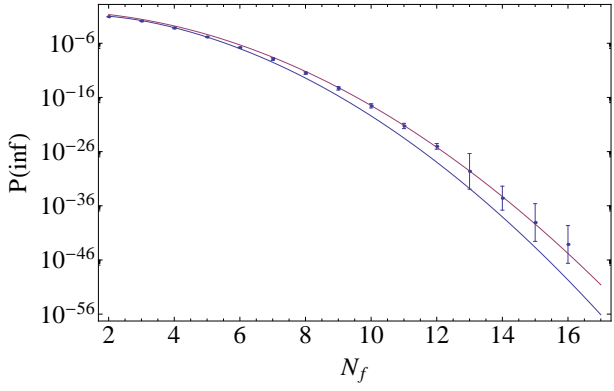} 
\caption{Probability for inflation as a function of $N_f$. In blue (lower line) the analytical estimate of Eq. (\ref{eq:Pinf}) and in red (upper line) the best fit of Table \ref{fitResults}. }
\label{fig:Pinf}
\end{figure}

In Fig. \ref{fig:Pinf} we plot the probability for finding an inflationary saddle point in the landscape, presenting both the data points, the 
analytical estimate \cite{LiamForwardQuote}
\be
P(inf)=e^{-2 \Phi(-\eta)N_f^2}-e^{-2 \Phi (\eta) N_f^2}.
\label{eq:Pinf}
\ee
and the best fit of Table \ref{fitResults}.

\begin{figure}[h!]
\centering
\includegraphics[width=0.45\textwidth]{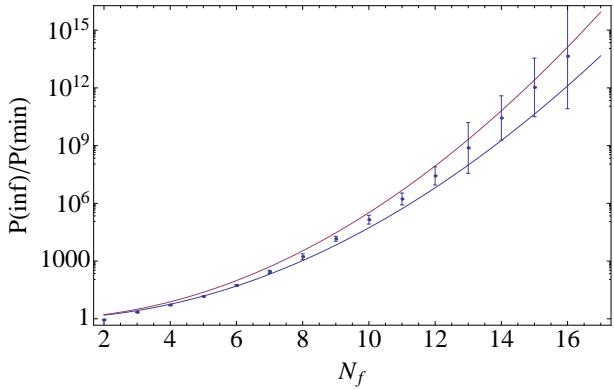}
\caption{$P(inf)/P(min)$: Inflationary saddle points are exponentially more likely than local minima in the Wigner landscape. In blue (lower line) the analytical estimate and in red (upper line) the best fit. }
\label{fig:Inf2Min}
\end{figure}

As anticipated flat saddle points, like minima, are extremely unlikely in the Wigner landscape as they correspond to large fluctuations of the smallest eigenvalue. However since it is statistically costlier to displace the smallest eigenvalue to $0$ than to $-\eta=-0.1$, flat saddle points are exponentially more abundant than local minima  as is illustrated in Fig. \ref{fig:Inf2Min}. The ratio given by
\be
\frac{P(inf)}{P(min)}\sim \left\{
 \begin{array}{rl}
  e^{ 0.127 N_f^2} & \text{fitted} \\
  e^{ 0.109 N_f^2} & \text{analytical }
 \end{array} \right. .
\ee

We will now relate this behaviour in terms of the cutoff $\eta$ on the mass of the fields to the 2nd slow-roll condition $\eta_V=\eta/V<1$. For this purpose, we note that our results above were obtained by choosing the variance of the Wigner ensemble to be $\sigma=1/\sqrt{2N_f}$. This approximates a random supergravity where the mass eigenvalues distribute according to the Wigner semi-circle law on a range $[-\sqrt 2, \sqrt 2]$ in units of $M_P$. The crucial point to observe is that a typical supergravity landscape has both its typical potential energy and mass eigenvalue scale characterised by the gravitino mass $m_{3/2}=e^{K/2}W$ as this controls the typical size of the individual contributions in \eqref{eq:VF}: $|\langle V\rangle|\sim m_{3/2}^2\sim \sqrt{\langle(\partial_i\partial_j V)^2\rangle}$. Therefore, the choice $\sigma=1/\sqrt{2 N_f}$ with its typical mass eigenvalue size of ${\cal O}(1)$ describes random supergravities with $m_{3/2}\sim {\cal O}(1)$. Since for such supergravities we then also have $|\langle V\rangle|\sim m_{3/2}^2\sim {\cal O}(1)$, we have $\eta\sim \eta_V$ and a cutoff $\eta <1$ in the integrations of \eqref{eq:pdfInt} directly implies slow-roll. The study of actual string theory derived example landscapes~\cite{Kachru:2003aw,Balasubramanian:2005zx,Balasubramanian:2004uy,Rummel:2011cd} points to scenarios where
$|\langle V\rangle|\sim m_{3/2}^2\lesssim M_{\rm GUT}^2\sim 10^{-5}$. We can now use the Wigner semi-circle law \eqref{eq:semicircle} together with the joint pdf \eqref{eq:jointpdf} to rescale $\sigma\to\sigma\, m_{3/2}^2$ which will approximate the mass eigenvalue distribution of a random supergravity with $|\langle V\rangle|\sim \sqrt{\langle(\partial_i\partial_j V)^2\rangle}\sim m_{3/2}^2$ and eigenvalue range $[-\sqrt 2 m_{3/2}, \sqrt 2 m_{3/2}]$. This forces us to rescale the integration limits in \eqref{eq:pdfInt} to $\pm \eta\, m_{3/2}^2$. As we now have $\sqrt{\langle(\partial_i\partial_j V)^2\rangle}\sim m_{3/2}^2$, we now get that the 2nd slow-roll parameter $\eta_V=\eta\, m_{3/2}^2 / \sqrt{\langle(\partial_i\partial_j V)^2\rangle}\sim \eta$ is again specified by the original cutoff $\eta <1$. Therefore, the exponential enhancement which we found above for $m_{3/2}\sim {\cal O}(1)$ generalises to the known string landscape regions which can be approximated by random supergravities with $m_{3/2}\lesssim M_{\rm GUT}$ controlling both the typical size of the scalar potential and the mass matrix eigenvalue size.

Note that this exponential enhancement is estimated conservatively, as the random matrix description of the critical points of a random supergravity by definition selects for either minima or saddle points. Yet, small-field inflationary regions do exist on almost flat inflection points of the scalar potential as well, with a tuning cost comparable to that of flat saddle point. Therefore, our method is conservative in that it underestimates the total rate of small-field inflationary regions occurring in a given random supergravity.

\begin{figure}[h]
\centering
\includegraphics[width=0.45\textwidth]{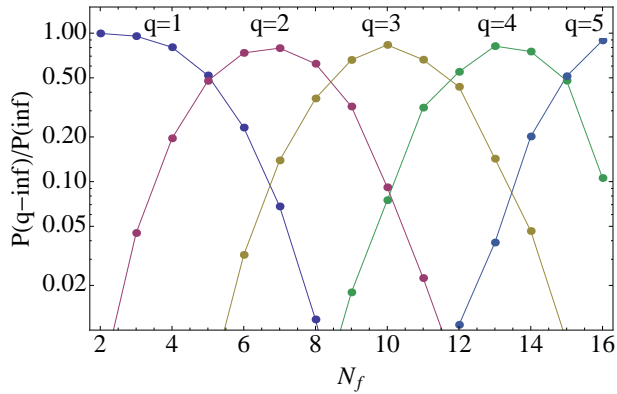}
\caption{Relative likelihood of q-field inflation as a function of the total dimensionality of the field space $N_f$.}
\label{fig:WhatInf}
\end{figure}

The same method that lead us to the above conclusions also allows us to discern what is the preferred inflationary dynamics for a given $N_f$. Dyson's interpretation of Eq. (\ref{eq:jointpdf}) in terms of a gas of charged particles gives us a hint of what behaviour to expect. For any particular value of $N_f$ there are $N_f$ possible types of inflationary dynamics: from single field to $N_f$ field inflation. Single field inflation corresponds to having only one eigenvalue in the range $[-\eta, \eta]$ and the remaining $N_f-1$ in $[\eta,\infty[$. For large values of $N_f$ this is highly unlikely since eigenvalue repulsion in the interval $[\eta,\infty[$ would tend to push one or more eigenvalues into the inflationary region.
On the other hand $N_f$ field inflation is also very rare, since it corresponds to squeezing all eigenvalues in the narrow range $[-\eta,\eta]$, leading to a configuration where the repulsive force would tend to push some eigenvalues out of this interval. Somewhere between these two limiting cases one can find the most likely behaviour. 
In Fig. \ref{fig:WhatInf} we plot the ratio $P(q-inf)/P(inf)$ as a function of $N_f$ for $\eta=0.1$.

We observe that the transition from single to two field inflation happens at $N_f=5$ with the next transitions from 2 to 3  and 3 to 4 field inflation happening around 8 and 12 respectively. We note that the values for which the various transitions happen depend strongly on $\eta$:  the larger the $\eta$ the sooner the transitions will happen. A quantitative understanding may be developed by studying the distribution of spacings between adjacent eigenvalues.


Next, we recall that the minimum total number of e-folds of slow-roll inflation at a critical point scales with $\eta$ as $N_{tot}\sim 1/|\eta|$~\cite{Freivogel:2005vv,LiamForwardQuote}. The question of whether we should select for the maximum amount of slow-roll inflation (due to the maximised 3-volume growth) or not amounts to a choice of the measure of eternal inflation. Therefore the answer to the question whether we expect single-small-field or multi-small-field inflation to dominate the small-field regime likely depends on the choice of the measure.

The presence of several fields contributing to inflation close to a saddle point or inflection point has the potential of generating local non-Gaussianity which is absent in the single-field case. As this is tied to the relative importance of single-field versus multi-field, statements about possible non-Gaussianity emanating from a multi-small-field regime again likely depend on the choice of the measure.  We leave this for future work.


\section{Conclusions}

In this note we have determine the number frequency count of small-field inflation models on the landscape of supersymmetric type IIB CY flux vacua. As the effective 4D theory of both dS vacua and small-field modular inflation models in this region of the landscape is described by 4D ${\cal N}=1$ supergravity, we have used random matrix theory to describe the region's vacuum structure in terms of a random supergravity~\cite{Denef:2004cf,Marsh:2011aa}. Meta-stable dS vacua require a fully positive-definite mass matrix (Hessian). Such Hessians constitute an exponentially suppressed fluctuation of all eigenvalues to positivity in the context of the theory of the random Hessians from 4D ${\cal N}=1$ supergravity. Consequently, we expected small-field inflation models which relax the positivity for at least one of the eigenvalues to be favoured compared to full meta-stability. Our analysis of the Wigner ensemble giving the leading order description of this effect in random supergravity matched this expectation. We find that there are exponentially many more small-field inflation models in the moduli potential of the type IIB flux landscape than there are proper dS vacua. The analysis of the frequency of large-field models and the cosmological probability factor in~\cite{Westphal:2012up} led to an estimate for the relative likelihood of large-field inflation
\be
\frac{P_{\Delta\phi_{60}>M_P}}{P_{\Delta\phi_{60}<M_P}}\sim\langle h^{1,1}_-\rangle\,\frac{\beta_{h^{1,1}_{-}\geq1}}{\beta_{flat\;saddle}}\quad.
\ee
We may now plug in that~\cite{Westphal:2012up} $\beta_{h^{1,1}_{-}\geq1}<1$ (not all CYs will support the topological requirements for axion monodromy) and $\langle h^{1,1}_-\rangle<h^{1,1}\lesssim{\cal O}(100)$, as well as our results here $\beta_{flat\;saddle}\sim \exp(+\delta c N_f^2)\gg 1$. Finally, we note that the small-field model enhancement $\beta_{flat\;saddle}$ is the largest for the least tuned saddle points with $|\eta|\sim{\cal O}(0.1)$. Upon imposing COBE normalisation on the generation of inflationary curvature perturbations, these saddle points also have the largest energy scales of all small-field models, and in turn only moderately sub-Planckian field ranges. Hence, we predict small-field inflation to dominate abundantly, and to be concentrated at the largest energy scales compatible with a sub-Planckian field range. Consequently, we expect a typical tensor-to-scalar ratio $r\sim {\cal O}(10^{-3})$ which may be within reach of future CMB B-mode polarisation measurements.

\section*{Acknowledgments}
We would like to thank D.~Marsh, L.~McAllister, E.~Pajer and T.~Wrase for useful comments on a earlier version of this work. This work was supported by the Impuls und Vernetzungsfond of the Helmholtz Association of German Research Centres under grant HZ-NG-603.


\begin{thebibliography}{10}

\bibitem{Story:2012wx} 
  K.~T.~Story, C.~L.~Reichardt, Z.~Hou, R.~Keisler, K.~A.~Aird, B.~A.~Benson, L.~E.~Bleem and J.~E.~Carlstrom {\it et al.},
  arXiv:1210.7231 [astro-ph.CO].
  
  \bibitem{Hinshaw:2012fq} 
  G.~Hinshaw, D.~Larson, E.~Komatsu, D.~N.~Spergel, C.~L.~Bennett, J.~Dunkley, M.~R.~Nolta and M.~Halpern {\it et al.},
  arXiv:1212.5226 [astro-ph.CO].
  
  \bibitem{Sievers:2013wk} 
  J.~L.~Sievers, R.~A.~Hlozek, M.~R.~Nolta, V.~Acquaviva, G.~E.~Addison, P.~A.~R.~Ade, P.~Aguirre and M.~Amiri {\it et al.},
  arXiv:1301.0824 [astro-ph.CO].
  
  \bibitem{Riess:2011yx} 
  A.~G.~Riess, L.~Macri, S.~Casertano, H.~Lampeitl, H.~C.~Ferguson, A.~V.~Filippenko, S.~W.~Jha and W.~Li {\it et al.},
  Astrophys.\ J.\  {\bf 730}, 119 (2011)
  [Erratum-ibid.\  {\bf 732}, 129 (2011)]
  [arXiv:1103.2976 [astro-ph.CO]].
  
  \bibitem{Riess:1998cb} 
  A.~G.~Riess {\it et al.}  [Supernova Search Team Collaboration],
  Astron.\ J.\  {\bf 116}, 1009 (1998)
  [astro-ph/9805201].
  
  \bibitem{Perlmutter:1998np} 
  S.~Perlmutter {\it et al.}  [Supernova Cosmology Project Collaboration],
  Astrophys.\ J.\  {\bf 517}, 565 (1999)
  [astro-ph/9812133].
  
  
  
  \bibitem{Blake:2011en} 
  C.~Blake, E.~Kazin, F.~Beutler, T.~Davis, D.~Parkinson, S.~Brough, M.~Colless and C.~Contreras {\it et al.},
  Mon.\ Not.\ Roy.\ Astron.\ Soc.\  {\bf 418}, 1707 (2011)
  [arXiv:1108.2635 [astro-ph.CO]].
  
  \bibitem{Padmanabhan:2012hf} 
  N.~Padmanabhan, X.~Xu, D.~J.~Eisenstein, R.~Scalzo, A.~J.~Cuesta, K.~T.~Mehta and E.~Kazin,
  arXiv:1202.0090 [astro-ph.CO].
  
  
  \bibitem{Anderson:2012sa} 
  L.~Anderson, E.~Aubourg, S.~Bailey, D.~Bizyaev, M.~Blanton, A.~S.~Bolton, J.~Brinkmann and J.~R.~Brownstein {\it et al.},
  Mon.\ Not.\ Roy.\ Astron.\ Soc.\  {\bf 428}, 1036 (2013)
  [arXiv:1203.6594 [astro-ph.CO]].
  
  




\bibitem{Guth:1980zm} 
  A.~H.~Guth,
  Phys.\ Rev.\ D {\bf 23}, 347 (1981).

  
  \bibitem{Linde:1981mu} 
  A.~D.~Linde,
  Phys.\ Lett.\ B {\bf 108}, 389 (1982).
  
  
  \bibitem{Albrecht:1982wi} 
  A.~Albrecht and P.~J.~Steinhardt,
  Phys.\ Rev.\ Lett.\  {\bf 48}, 1220 (1982).
  
  \bibitem{Linde:1983gd} 
  A.~D.~Linde,
  Phys.\ Lett.\ B {\bf 129}, 177 (1983).
  
  
\bibitem{Baumann:2009ds} 
  D.~Baumann,
  arXiv:0907.5424 [hep-th].
  
  \bibitem{Bousso:2000xa} 
  R.~Bousso and J.~Polchinski,
  JHEP {\bf 0006}, 006 (2000)
  [hep-th/0004134].
  
  \bibitem{Kachru:2003aw}
  S.~Kachru, R.~Kallosh, A.~D.~Linde and S.~P.~Trivedi,
  Phys.\ Rev.\ D {\bf 68} (2003) 046005
  [hep-th/0301240].


\bibitem{Susskind:2003kw} 
  L.~Susskind,
  In *Carr, Bernard (ed.): Universe or multiverse?* 247-266
  [hep-th/0302219].
  
\bibitem{Douglas:2003um} 
  M.~R.~Douglas,
  JHEP {\bf 0305}, 046 (2003)
  [hep-th/0303194].
  
\bibitem{Denef:2004cf}
  F.~Denef and M.~R.~Douglas,
  JHEP {\bf 0503} (2005) 061
  [hep-th/0411183].


  
\bibitem{Douglas:2006es} 
  M.~R.~Douglas and S.~Kachru,
  Rev.\ Mod.\ Phys.\  {\bf 79}, 733 (2007)
  [hep-th/0610102].
  
  
   \bibitem{Kachru:2003sx} 
  S.~Kachru, R.~Kallosh, A.~D.~Linde, J.~M.~Maldacena, L.~P.~McAllister and S.~P.~Trivedi,
  JCAP {\bf 0310}, 013 (2003)
  [hep-th/0308055].
  
  
  \bibitem{Baumann:2009ni} 
  D.~Baumann and L.~McAllister,
  Ann.\ Rev.\ Nucl.\ Part.\ Sci.\  {\bf 59}, 67 (2009)
  [arXiv:0901.0265 [hep-th]].
  
  \bibitem{Cicoli:2011zz} 
  M.~Cicoli and F.~Quevedo,
  Class.\ Quant.\ Grav.\  {\bf 28}, 204001 (2011)
  [arXiv:1108.2659 [hep-th]].

  
  
  
  
  \bibitem{Freese:1990rb} 
  K.~Freese, J.~A.~Frieman and A.~V.~Olinto,
  Phys.\ Rev.\ Lett.\  {\bf 65}, 3233 (1990).
  
  \bibitem{Banks:2003sx} 
  T.~Banks, M.~Dine, P.~J.~Fox and E.~Gorbatov,
  JCAP {\bf 0306}, 001 (2003)
  [hep-th/0303252].
  
  \bibitem{Dimopoulos:2005ac} 
  S.~Dimopoulos, S.~Kachru, J.~McGreevy and J.~G.~Wacker,
  JCAP {\bf 0808}, 003 (2008)
  [hep-th/0507205].
  
  \bibitem{Svrcek:2006yi} 
  P.~Svrcek and E.~Witten,
  JHEP {\bf 0606}, 051 (2006)
  [hep-th/0605206].
  
    \bibitem{McAllister:2008hb} 
  L.~McAllister, E.~Silverstein and A.~Westphal,
  Phys.\ Rev.\ D {\bf 82}, 046003 (2010)
  [arXiv:0808.0706 [hep-th]].
  
    \bibitem{Lyth:1996im}
  D.~H.~Lyth,
  Phys.\ Rev.\ Lett.\  {\bf 78} (1997) 1861
  [hep-ph/9606387].





 
  

  
  
    
  
  
  
  
  \bibitem{BenDayan:2009kv} 
  I.~Ben-Dayan and R.~Brustein,
  JCAP {\bf 1009}, 007 (2010)
  [arXiv:0907.2384 [astro-ph.CO]].
  

  
\bibitem{Hotchkiss:2011gz} 
  S.~Hotchkiss, A.~Mazumdar and S.~Nadathur,
  JCAP {\bf 1202}, 008 (2012)
  [arXiv:1110.5389 [astro-ph.CO]].
  
  
  \bibitem{Barnaby:2012xt} 
  N.~Barnaby, J.~Moxon, R.~Namba, M.~Peloso, G.~Shiu and P.~Zhou,
  Phys.\ Rev.\ D {\bf 86}, 103508 (2012)
  [arXiv:1206.6117 [astro-ph.CO]].
  
 \bibitem{Senatore:2011sp} 
  L.~Senatore, E.~Silverstein and M.~Zaldarriaga,
  arXiv:1109.0542 [hep-th].
\bibitem{Kobayashi:2013awa} 
  T.~Kobayashi and T.~Takahashi,
  arXiv:1303.0242 [astro-ph.CO].



  
  



  \bibitem{Weinberg:1987dv} 
  S.~Weinberg,
  Phys.\ Rev.\ Lett.\  {\bf 59}, 2607 (1987).
  
  
   \bibitem{Coleman:1980aw} 
  S.~R.~Coleman and F.~De Luccia,
  Phys.\ Rev.\ D {\bf 21}, 3305 (1980).


\bibitem{Westphal:2012up} 
  A.~Westphal,
  arXiv:1206.4034 [hep-th].
  
  
  



\bibitem{Marsh:2011aa}
  D.~Marsh, L.~McAllister and T.~Wrase,
  JHEP {\bf 1203} (2012) 102
  [arXiv:1112.3034 [hep-th]].
  

\bibitem{Aazami:2005jf}
  A.~Aazami and R.~Easther,
  JCAP {\bf 0603} (2006) 013
  [hep-th/0512050].

\bibitem{Wishart:1928}
 J. Wishart, 
 Biometrika 20A, 32 (1928).

\bibitem{Metha:1967}
M. L. Mehta, ``Random Matrices and the statistical theory of energy levels'', Academic Press, 1967.

\bibitem{Guhr:1997ve}
  T.~Guhr, A.~Muller-Groeling and H.~A.~Weidenmuller,
  Phys.\ Rept.\  {\bf 299} (1998) 189
  [cond-mat/9707301].



\bibitem{Edelman:2005}
  A.~Edelman, N.~R.~Rao, Acta Numerica, 2005, 1-65
  
  \bibitem{Marchenko:1967}
V. A. Marchenko,  L. A. Pastur 
Mat. Sb. (N.S.), 72(114):4,(1967)  507Ð536 
  
\bibitem{Dyson:1962es}
  F.~J.~Dyson,
  J.\ Math.\ Phys.\  {\bf 3} (1962) 140.

  
\bibitem{Dean:2006wk}
  D.~S.~Dean and S.~N.~Majumdar,
  Phys.\ Rev.\ Lett.\  {\bf 97} (2006) 160201
  [cond-mat/0609651].\\
  D. S. Dean and S. N. Majumdar, 
  Phys. Rev. E 77, 041108 (2008), arXiv:0801.1730v1 [cond-mat.stat- mech].


\bibitem{LiamForwardQuote}
 D.~Marsh,  L.~McAllister,  E.~Pajer \& T.~Wrase,
  [arXiv:1203.xxxx].
  

\bibitem{Balasubramanian:2005zx} 
  V.~Balasubramanian, P.~Berglund, J.~P.~Conlon and F.~Quevedo,
  JHEP {\bf 0503}, 007 (2005)
  [hep-th/0502058].
  
  \bibitem{Balasubramanian:2004uy} 
  V.~Balasubramanian and P.~Berglund,
  JHEP {\bf 0411}, 085 (2004)
  [hep-th/0408054].
  
  \bibitem{Rummel:2011cd} 
  M.~Rummel and A.~Westphal,
  JHEP {\bf 1201}, 020 (2012)
  [arXiv:1107.2115 [hep-th]].
  

\bibitem{Freivogel:2005vv} 
  B.~Freivogel, M.~Kleban, M.~Rodriguez Martinez and L.~Susskind,
JHEP {\bf 0603}, 039 (2006)
[hep-th/0505232].



  \end{thebibliography}
\end{document}